# Electric field tunable exchange interaction in InAs/GaAs coupled quantum dots

Kushal C. Wijesundara,[1] Mauricio Garrido,[1] Swati Ramanathan,[1] E. A. Stinaff,[1]
M. Scheibner,[2] A. S. Bracker,[2] and D. Gammon[2]

[1] Department of Physics and Astronomy, and Nanoscale and Quantum Phenomena Institute,
  Ohio University, Athens, OH 45701-2979, USA.
[2] Naval Research Laboratory, Washington, DC 20375, USA.

**ABSTRACT**

Spin manipulation in coupled quantum dots is of interest for quantum information applications. Control of the exchange interaction between electrons and holes via an applied electric field may provide a promising technique for such spin control. Polarization dependent photoluminescence (PL) spectra were used to investigate the spin dependent interactions in coupled quantum dot systems and by varying an electric field, the ground state hole energy levels are brought into resonance, resulting in the formation of molecular orbitals observed as anticrossings between the direct and indirect transitions in the spectra. The indirect and direct transitions of the neutral exciton demonstrate high and low circular polarization memory respectively due to variation in the exchange interaction. The ratio between the polarization values as a function of electric field, and the barrier height was measured. These results indicate a possible method of tuning between indirect and direct configurations to control the degree of exchange interaction.

**INTRODUCTION**

Quantum computational algorithms were developed decades ago but practical implementation of such a device is yet to be achieved, although considerable efforts have been devoted to realize the next generation quantum information processing and quantum computing schemes[1]. Of the possible platforms on which to implement these algorithms, spins in coupled quantum dots (CQDs) are thought to be excellent candidates for qubits as they can be initialized, manipulated and measured through optical spectroscopy. Progress has been made through the use of single quantum dots (SQDs) in implementing optical quantum gates [2], ultra fast control of spin excitons [3], and entanglement of the electron-hole complexes [4]. The need for scalability has led to increasing interest in CQDs where one may have new options in controlling the spin and charge states along with their coupling and interactions [5]. Encoding and manipulation of spins have already been demonstrated in different solid state systems, such as the phenomena of quantum interference have been utilized for spin manipulation in SQDs with coherent micro spectroscopy techniques [6]. Furthermore coherent manipulation of spins has been demonstrated in quantum two level systems [7].

In this paper we investigate, through optical spectroscopy, the control of exchange interaction by spatially separating the electron and hole in a CQD system. Measurement of the isotropic electron-hole exchange splitting between the bright and dark exciton can be performed in CQD systems without the use of an applied magnetic field [8]. However, the anisotropic exchange splitting tends to be small and difficult to measure (even in the intradot exciton). It may however be possible, through polarization measurements, to gain insight into the details of the anisotropic exchange interaction as the CQD is selectively tuned with electric field.

In general the magnitude of the exchange interaction depends on the overall spin state along with the spatial overlap of the electrons and holes. Therefore when both electron and hole are in the bottom dot the peaks of the amplitudes of the wave functions gives rise to maximum overlaps hence, maximum exchange interaction. But when the electron and hole are spatially separated and reside in top and bottom dots, as in the case of an indirect exciton, the overlap of the wave functions reduces giving rise to a lower exchange interaction which tends to zero as we increasingly separate the electron and the hole.

**EXPERIMENT**

Self-assembled InAs/GaAs quantum dots grown in the S-K mode were used in this study. To control the relative energy levels of the two dots and selectively charge them, the CQDs were embedded in a Schottky diode structure with an n-doped back contact and a semitransparent Ti top contact which allows the application of an electric field along the growth direction. The asymmetric nature of the CQDs gives rise to energy level resonances under desired bias voltages which is a vital step in implementing spin manipulation for quantum information process [9,10]. Further, lithographically defined submicron apertures are created with aluminum mask to optically probe individual quantum dots. Samples were kept at cryogenic temperatures and were excited with a Ti: Sapphire laser operated in CW mode. The detection is made using a triple grating spectrometer coupled to a liquid-nitrogen cooled CCD detector.

Mapping the charge and spin states is done using bias maps which are three dimensional contour plots with applied electric field, PL spectral energy, and PL intensity represented by 'x', 'y', and 'z' coordinates respectively. Similar to previously studied single dot bias maps [11], in CQDs we recognize strong PL lines that show little electric field dependence and are identified as various intra-dot charged exciton states. Relative to the neutral exciton ($X^0$), the positively charged exciton (positive trion $X^+$) is found to be typically ~2 meV higher in energy, whereas the negatively charged exciton (negative trion $X^-$) is ~6 meV lower and new features such as large E-field dependence, anticrossings, and 'X' pattern can be identified in CQD spectra [10]. These new features arise from tunnel coupling of the hole states resulting in the formation of molecular states. This leads to anticrossings observed in the PL spectra as a function of applied electric field and are schematically shown in Figure 1 for $X^0$. At high electric fields, the hole levels become resonant leading to enhancement of the symmetric and anti symmetric molecular hole wave functions. Away from the anticrossing region the holes can be predominantly localized on either the top or bottom dot leading to indirect or direct excitonic states respectively (Figure 1). We represent the state of the CQD using $\begin{pmatrix} e_B & e_T \\ \underline{h_B} & h_T \end{pmatrix}$, where the four terms denote the number of electrons ($e$) and holes ($h$) in the top ($T$) and bottom ($B$). The underlined terms indicate which particles contribute to the recombination. By using arrows we may also denote the spin state of the charges, $\begin{pmatrix} \uparrow_B & \downarrow_T \\ \underline{\Downarrow_B} & \Uparrow_T \end{pmatrix}$ with spin up or down electrons (holes) in the bottom dot $\uparrow_B, \downarrow_B$ ($\Downarrow_B, \Uparrow_B$) and in the top dot $\uparrow_T, \downarrow_T$ ($\Downarrow_T, \Uparrow_T$).

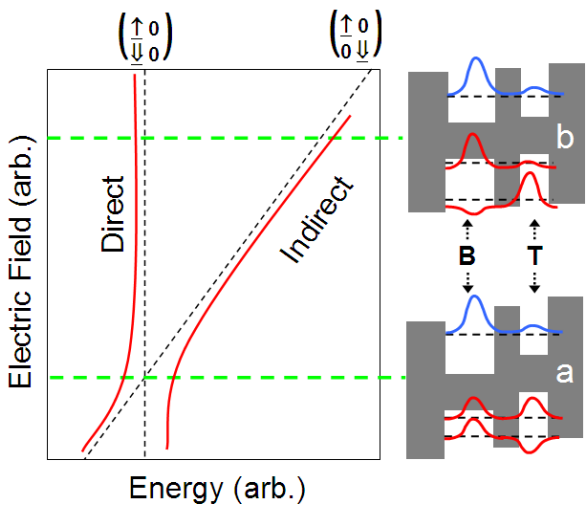

**Figure 1.** (a) Schematic representation of the CQDs in a Schottky diode structure with larger bottom dot (B) and smaller top dot (T) at high enough electric fields shows hole level resonances along with electron wave function (blue) symmetric (direct) and anti symmetric (indirect) hole wave functions (red). (b) At reverse electric fields lesser overlap between the electron and the indirect hole wave function is shown. On the left a schematic representation of the anticrossing signature of the direct and indirect exciton with their relevant spin configurations.

We use polarization dependent PL to help identify the spin states and interactions associated with different exitonic states in the CQD system. Circularly polarized laser excitation is used to couple to the spin states of the carriers as there is a relationship between photon polarization and electron, hole spin projection due to angular momentum conservation [12]. Polarization dependent photoluminescence is accomplished by introducing linear polarizers and liquid crystal retarders to the basic PL experiment. Initially linear polarized light is created by sending the CW laser beam with 860 nm through the linear polarizer and the desired circular polarized light is generated using the liquid crystal retarder which excites the CQDs. By varying the applied voltage given to the liquid crystal retarder, its retardance can be varied which is used to generate a phase shift of λ/4 or 3λ/4 to create σ⁺ or σ⁻ circularly polarized light. After the recombination process, to analyze the degree of circular polarization, the emitted light is sent through a second liquid crystal retarder to detect both σ⁺ and σ⁻.

The degree of polarization is given by $P = (I^+ - I^-)/(I^+ + I^-)$ with $I^+(I^-)$ representing intensity of σ⁺ (σ⁻) polarization under σ⁺ polarized excitation. In the absence of exchange interaction we would anticipate 100 % degree of circular polarization, but in reality we observe different polarization memory values associated with different charge and spin configurations of the excitons which is a direct result of the exchange interaction.

**RESULTS AND DISCUSSION**

Polarization dependent photoluminescence spectra for the three different charge states reveal that the neutral exciton has vanishingly small circular polarization memory, while the positive trion and the negative trion retain much of their circularly polarization memory. Similar results have been observed in SQDs and understood as arising from the anisotropic exchange interaction [13,14]. In the case of neutral exciton, the electron and a hole experience a strong anisotropic exchange interaction resulting in the circular polarization memory being wiped out. Whereas, in the case of the positive trion which has two holes and one electron, the ground state hole spins must form a spin singlet resulting in a zero total spin, cancelling out the exchange interaction. Therefore the unpaired electron spin uniquely determines the polarization of the emission which is observed to be positive implying that the electron spin has largely retained its initial spin. Similarly for the negative trion with two paired electrons and a hole, the degree of polarization is determined by the spin of the hole since the total electron spin is zero.

In the previous discussion it was shown that the effects of the anisotropic exchange interaction can be mitigated by the elimination of the overall spin. In the CQD system we also have the ability to spatially separate the electron and hole, providing another way to affect the overall exchange interaction. In CQDs we can tune the excitonic emission from intra-dot to inter-dot by simply tuning the electric field. As seen in Figure 1, by varying the electric field one can tune from indirect to direct transition energy. This has the effect of shifting the wave function amplitude of the hole and therefore changing the overall electron-hole wave function overlap. From polarization dependent spectra (Figure 2), the direct transition demonstrates a weaker degree of circular polarization memory as it shows strong exchange interaction in its direct configuration, whereas the indirect transition shows a relatively higher degree of circular polarization memory due to the reduced exchange interaction. Furthermore at the anticrossing regime which happens around the applied electric field of 53.96 kV/cm, the degree of polarization for both indirect and direct transitions tend to be almost the same. This is because our n-doped samples demonstrate hole level resonances and at this electric field the molecular symmetric and anti-symmetric hole wave functions have comparable amplitude in both the top and bottom dots. This results in an increase in the polarization memory of the direct configuration of the neutral exciton and a reduction in the polarization memory in its indirect configuration. These results are experimentally observed for a 4nm barrier CQD sample as shown in Figure 2 with the indirect line shown by solid squares, and direct line by open squares. Since the anisotropic exchange interaction directly affects the degree of circular polarization memory, by observing the variation in the degree of polarization we deduce that by tuning the electric field we are able to change the overall electron-hole exchange interaction.

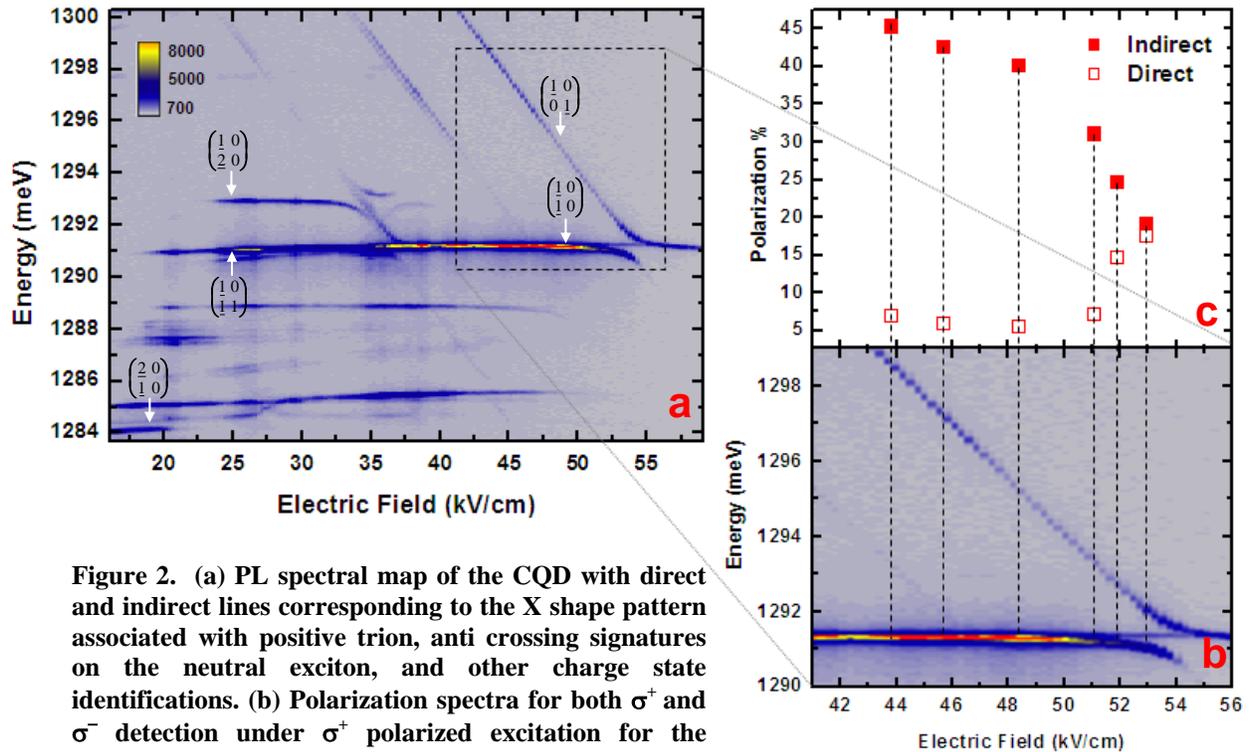

**Figure 2.** (a) PL spectral map of the CQD with direct and indirect lines corresponding to the X shape pattern associated with positive trion, anti crossing signatures on the neutral exciton, and other charge state identifications. (b) Polarization spectra for both $\sigma^+$ and $\sigma^-$ detection under $\sigma^+$ polarized excitation for the neutral exciton near anticrossing.
(c) Measured polarization memory corresponding to specific electric field for direct (open squares) and indirect (solid squares). These experiments were performed using an n-doped coupled dot structure with a 4 nm barrier between the top and bottom dot with CW excitation of 913 nm.

We also performed barrier dependent experiments using similar CQD structures with barrier heights of 4 nm and 2 nm between the top and bottom dot respectively. All the experimental parameters were kept constant for both samples in polarization memory measurements. Due to variations between CQDs we typically observe hole level anti crossing at different electric fields. For the 2nm barrier CQDs we found the field at which the anticrossing occurred to be 43.38 kV/cm, and 53.96 kV/cm for the 4 nm barrier. To compare the two samples we plot the polarization measurements at electric fields relative to the anticrossing point.

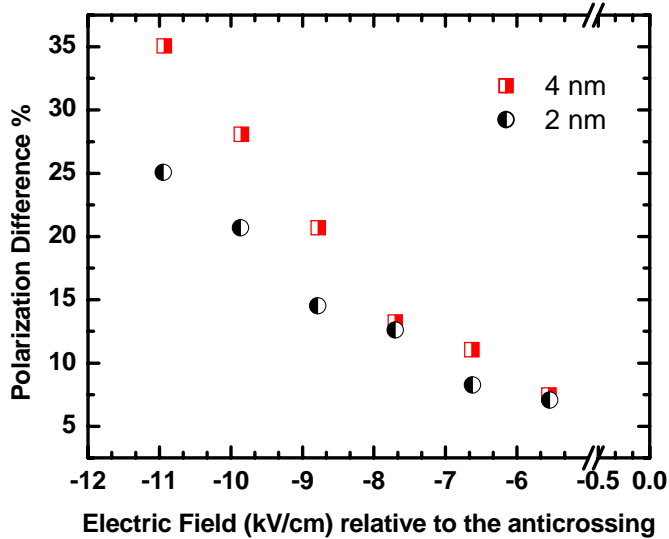

**Figure 3.** Difference of the degree of circular polarization for the indirect and direct configuration of the neutral exciton with 4nm barrier (squares) and 2nm barrier (circles) as shown here. This trend in the degree of polarization which is a measure of the exchange interaction as a function of electric field shows a common trend with barrier heights.

Figure 3 shows the difference in the degree of polarization for the indirect and direct transitions of the neutral exciton for both 2 nm and 4 nm as a function of electric field relative to the anticrossing. In general we observe an increase in the polarization memory as the electric field is lowered with respect to the anticrossing. Further we observe that as the barrier increases the degree of polarization increases as can be seen from the squares corresponding to 4 nm barrier compared to the circles corresponding to the 2 nm barrier. This is due to the fact that as the barrier between the top and the bottom dot is increased, the degree of overlap of the hole wave function with the electron wave function reduces.

**CONCLUSIONS**

The variation of the degree of polarization has been studied in the ground state of a neutral exciton in the InAs/GaAs CQD structure. Spectra from the direct and indirect configurations of the neutral exciton were analyzed for both circular polarizations. The low degree of circular polarization memory in the direct configuration of the neutral exciton is a consequence of the anisotropic exchange interaction. At lower electric fields the electron and the hole tend to spatially separate resulting in a reduction of the wave function overlap and hence the exchange interaction, which is observed as an increase in circular polarization memory. Further the increase in barrier separation results in an enhancement of this effect. Therefore, for CQDs, an applied electric field can be used to tune between indirect and direct configurations to control the degree of exchange interaction, providing a possible tool for spin manipulation.

**ACKNOWLEDGMENTS**

We would like to thank CMSS and NQPI for their financial support.